\documentclass[aps,preprint]{revtex4}%
\usepackage{amssymb}
\usepackage{amsfonts}
\usepackage{amsmath}
\usepackage{graphicx}%
\providecommand{\U}[1]{\protect\rule{.1in}{.1in}}

\begin{document}
\title[Short title for running header]{Non-perturbative Solution to the Quantum Interaction Problem via Schwinger's
Action Principle}
\author{C. A. M. de Melo$^{1,2}\footnote{cassius@unifal-mg.br}$}
\author{B. M. Pimentel$^{2}\footnote{pimentel@ift.unesp.br}$}
\author{J.A. Ramirez$^{3}\footnote{johnarb@ufba.br}$}
\affiliation{$^{1}$Instituto de Ci\^{e}ncia e Tecnologia - Universidade Federal de Alfenas,
Campus Po\c{c}os de Caldas, BR 267 - Rod. Pref. Jos\'{e} Aur\'{e}lio Vilela
11.999, Km 533, 37715-400 Cidade Universit\'{a}ria - Po\c{c}os de Caldas, MG, Brasil}
\affiliation{$^{2}$Instituto de F\'{\i}sica Te\'{o}rica - S\~{a}o Paulo State University,
P.O.Box 70532-2, CEP: 01156-970 - S\~{a}o Paulo, SP, Brasil.}
\affiliation{$^{3}$Instituto de F\'{\i}sica - Universidade Federal da Bahia, Campus
Universit\'{a}rio de Ondina, CEP 40210-340, Salvador, BA, Brazil.}
\pacs{PACS number}

\begin{abstract}
The most realistic situations in quantum mechanics involve the interaction
between two or more systems. In the most of reliable models, the form and
structure of the interactions generate differential equations which are, in
the most of cases, almost impossible to solve exactly. In this paper, using
the Schwinger Quantum Action Principle, we found the time transformation
function that solves exactly the harmonic oscillator interacting with a set of
other harmonic coupled oscillators. In order to do it, we have introduced a new
special set of creation and annihilation operators which leads directly to the
\emph{dressed states} associated to the system, which are the real quantum
states of the interacting \emph{\textquotedblleft
field-particle\textquotedblright} system. To obtain the closed solution, it is
introduced in the same foot a set of \emph{normal mode} creation and
annihilation operators of the system related to the first ones by an
orthogonal transformation. We find the eigenstates, amplitude transitions and
the system spectra without any approximation. At last, we show that the
Schwinger Variational Principle provides the solutions in a free
representation way.

\end{abstract}
\volumeyear{year}
\volumenumber{number}
\issuenumber{number}
\eid{identifier}
\date[Date text]{date}
\received[Received text]{date}

\revised[Revised text]{date}

\accepted[Accepted text]{date}

\published[Published text]{date}

\maketitle

\section{Introduction}

To obtain an exact solution to the radiation-matter interaction problem results in
a difficult or almost impossible task for several systems. In many cases
theoretical approaches based on the intensity or asymptotic behavior of the
system allow perturbative or computational treatments, giving an idea of the
functional form of the solutions. Despite this, those situations are not
general and are related to very specific laboratory conditions. On the other
hand, systems with strong interaction escape from those approaches. In
particular situations these system can be reduced to special simple cases as
the semi-classical\emph{ Rabi} \emph{Model} (RM) and its quantum version, the
so-called \emph{Jaynes-Cummings Model} (JCM) \cite{jaynes,jayncumm}. These
idealizations of the very original problem are still rich in phenomenology,
admitting nonperturbative approaches to cavity confined systems: a single atom
interacting with one monochromatic mode of the field in the cavity, composing
a subfield of quantum optics called Cavity Quantum Electrodynamics (CQED).

There is a generalized approach to CQED called \emph{Dressed Atom} which can
deal with systems that can not be treated by perturbative methods and leading
to analytical results, being first proposed by Cohen Tannoudji, Serge Haroche
and Nicole Polonsky \cite{cohen,polonsky,haroche}. In such approach, the atoms
or molecules into the cavity interact with the surrounding photons forming a
weakly interacting subsystem with the rest of the field; thus, the atom is
\emph{\textquotedblleft dressed\textquotedblright} by the photons of the
surrounding electromagnetic field \cite{CohenHaroche}. This description works
since only one of the field modes inside the cavity, or a superposition of
them, has photons with the necessary frequency to induce the transitions in
the atoms, then we can say that those photons dress the atom and the other
modes are empty and make the role of reservoir. Thus, radiation processes in
cavities can be put in a convenient base allowing to deal with some cases that
otherwise might not be addressed by perturbative methods.

However, one of the limitations of the Dressed State theory is given by the
loss of generality when the interactions have a non-linear character; since in
those cases the determination of such states loses its rigor
\cite{apcmalbuisson}. Alternatively, one can define Dressed or Renormalized
Coordinates \cite{andion,flores,flores1}, offering an extension of the Dressed
State concept which proves to be suitable to deal with systems where the
interactions between the atom and modes of the field are linear. Some
applications of this approach are the description of ohmic dissipation
\cite{flores3} \ and the radiation problem of a dipolar charge distribution
\cite{andion,flores,flores1,Pimen1,Pimen2}. It can be seen that, regardless of
the approach taken, either quantum mechanics \cite{cohen} or path integral
formalism \cite{Pimen1,Pimen2}, those studies always have been accompanied by
the need for representation of the system into a set of spatial coordinates.

Here we introduce a new approach to deal with systems modeled by a harmonic
oscillator interacting linearly with a set of other harmonic oscillators using
the Schwinger Quantum Action Principle. Instead of using Dressed Atom or
Coordinates approaches, we define two sets of creation and annihilation
operators for the real quantum states of each of the component of the system.
The first set will be associated with measurable physical states of each
individual component in the presence of interaction, \textit{i.e.}, to the
harmonic oscillator and each mode of the field. The second set is associated
with normal modes of the system, each of these modes expresses the state of the
whole system, oscillator + field modes. Thus, we shall call the first set
\emph{the creation and annihilation operators of Dressed States} of the system
(DSO) and the second set \emph{the creation and annihilation operators of
Normal Modes States} of the system (NMO).

The content of this work is divided into following: in the first section is
given a complete description of the model and a mathematical apart about the
characteristics of the interaction matrix. The second section is devoted to
define the two sets of the creation and annihilation operators as well as to
introduce the Normal and Dressed Hamiltonians. In the third section we briefly
introduce the Schwinger Action Principle and show how to obtain the dynamical
equations for the operators and the transformation function of the system. In
the last sections we find the final transformation function which is
calculated in the Normal Mode Operator base, and then it is transformed to the
Dressed State operator base from which is derived the spectrum and generalized
probability transitions. Finally, the work is ended with some observations and perspectives.

\section{Schwinger Quantum Action Principle}

Schwinger Quantum Action Principle offers a general approach to Quantum
Mechanics. This approach describes the microscopic phenomena by means of
quantum measurement symbols and expressing the dynamics of operators and
quantum states through a variational principle.

Since its foundation in the algebraic measurement theory
\cite{schwingercd,TeorMed} this quantum formalism emerges losing any reference
to the classical analogue of the system, not being a quantization procedure. In
this formulation, changes of quantum states and observables are studied in a
detailed way while the correspondence principle \cite{Diracbook} is not used
\emph{a priori}. Commutation relations are derived in a totally
self-consistent way \cite{Prinipvar,Kibble} from the analysis of the admisible
variations of the operators and states of the theory. Due to its generality,
Schwinger's formulation can be used to study quantization of gauge theories
without gauge fixing \cite{Bfield}, the behaviour of classical fields on
curved and torsioned backgrounds \cite{Lyra} and even to construct
quaternionic extensions of quantum theory \cite{Quaternions}.

In Schwinger's formulation any infinitesimal variation of a transformation
function $\left\langle a\left(  t_{1}\right)  |b\left(  t_{0}\right)
\right\rangle $ can be obtained as the matrix element of a single
infinitesimal generator: \emph{the infinitesimal quantum action operator}
\cite{schwingercd,Prinipvar},
\begin{align}
\delta\left\langle a\left(  t_{1}\right)  |b\left(  t_{0}\right)
\right\rangle  &  =i\left\langle a\left(  t_{1}\right)  \right\vert
\delta\widehat{S}_{t_{1},t_{0}}\left\vert b\left(  t_{0}\right)  \right\rangle
\nonumber\\
&  =i\left\langle a\left(  t_{1}\right)  \right\vert \left.  \left(  \hat
{p}\delta\hat{q}-\widehat{H}\delta t\right)  \right\vert _{t_{0}}^{t_{1}%
}\left\vert b\left(  t_{0}\right)  \right\rangle , \label{eq:actprinc}%
\end{align}
which is the variation $\delta\widehat{S}_{t_{1},t_{0}}=\delta\left[
\widehat{S}_{t_{1},t_{0}}\right]  $ of que quantum action operator
$\widehat{S}_{t_{1},t_{0}}=\int_{t_{0}}^{t_{1}}\widehat{L}(t)dt$. The
dynamical equations of Schr\"{o}dinger and Heisenberg can be obtained from
\eqref{eq:actprinc} fixing the boundary conditions whether on the states or on
the operators.

One of the essential concepts in this formalism is the fact that the dynamics
do not depend on the path of the system in the physical space, being all the
necessary information contained on the boundaries. This dependence is given by
the infinitesimal generator $\widehat{G}\left(  t\right)  $ evaluated in the
end time $t_{0}$ and $t_{1}$,
\begin{equation}
\widehat{G}=\hat{p}\delta\hat{q}-\widehat{H}\delta t, \label{eq:generator}%
\end{equation}
where%
\begin{gather}
\hat{p}\equiv\frac{\partial\widehat{L}}{\partial\dot{\hat{q}}}%
,\label{eq:canonicalform}\\
\hat{p}\cdot\dot{\hat{q}}-\widehat{L}\equiv\widehat{H}.\nonumber
\end{gather}

In order to integrate (\ref{eq:actprinc}) to obtain the probability amplitude,
it is necessary doing the time-ordering of the operators in the Dirac sense
\cite{Diracbook,Diraccd1} to reach $\delta\widehat{S}_{t_{1},t_{0}%
}\Longrightarrow\delta\widehat{\mathcal{W}}_{t_{1},t_{0}}$ such that
\begin{align*}
\delta\langle a(t_{1})|b(t_{0})\rangle &  =i\langle a(t_{1})|\delta
\widehat{S}_{t_{1},t_{0}}|b(t_{0})\rangle\\
&  =i\delta\mathcal{W}_{t_{1},t_{0}}\langle a(t_{1})|b(t_{0})\rangle
\end{align*}
obtaining,
\begin{equation}
\langle a(t_{1})|b(t_{0})\rangle=e^{i\mathcal{W}_{t_{1},t_{0}}}.
\label{amplitude}%
\end{equation}

Originally, the inclusion of the Lagrangian for the study of the dynamics at
quantum level was made by Dirac \cite{Diraccd1} arguing, among other things,
in favor of its relativistic invariance. We can see that the final form of the
Action Principle backs up to this idea, but Schwinger use a more convenient
form of the Lagrangian in its canonical form \eqref{eq:canonicalform},
\emph{i.e.,} as a function of momentum and coordinates, avoiding second
order differential equations for the observables.

This principle is particularly rich in analogies with classical mechanics,
since the generator $\hat{G}$, which is obtained from the variation of the
action and associated with all admissible infinitesimal variations of the
dynamic system, is also related to the unitary transformations which preserve
both the structure and characteristics of the studied system.

\section{A Useful Model}

In this work we shall use a model in which an harmonic oscillator is linearly
coupled to a set of many other harmonic oscillators. Such model appears in
several contexts in physics, specially when dealing with light-matter interaction.

In Ref. \cite{andion} this model is used to describe a harmonic oscillator in
the center of a sphere interacting with a scalar field. This model is a
simplification of a radiation-matter interaction problem in which an electric
dipole interacts with a electromagnetic field in the center of a spherical
cavity. In this case, let $q_{0}(t)$ be the position of the oscillator
representing the matter content and $\phi(\mathbf{r},t)$ the scalar field, the
equations of motion can be defined directly as
\begin{subequations}
\begin{align}
\ddot{q}_{0}(t)+\omega_{0}^{2}q_{0}(t)  &  =2\pi\sqrt{g}\int_{0}^{R}%
d^{3}\mathbf{r}\phi(\mathbf{r},t)\delta(\mathbf{r}),\label{eq:1}\\
\frac{\partial^{2}\phi(\mathbf{r},t)}{\partial t^{2}}-\bigtriangledown^{2}%
\phi(\mathbf{r},t)  &  =2\pi\sqrt{g}q_{0}(t)\delta(\mathbf{r}), \label{eq:2}%
\end{align}
in which $g$ is the interaction constant between the scalar field and the
harmonic oscillator, the light speed $c=1$ and $R$ is the sphere radius. The
boundary conditions on the radial coordinate $\mathbf{r}$ demands that
$\phi(\mathbf{r},t)$ be nonzero only for $0<\mathbf{r}<R$. The solution for
\eqref{eq:1} and \eqref{eq:2} can be expressed by performing an expansion
of the scalar field as follows
\end{subequations}
\begin{equation}
\phi\left(  \mathbf{r},t\right)  =\sum_{k=1}^{\infty}q_{k}\left(  t\right)
\phi_{k}\left(  \mathbf{r}\right)  ,
\end{equation}
in which, given the geometry of the system, an expansion in terms of plane
waves it is not appropriate. So, instead, we use an appropriate expansion of
the scalar field in terms of the resonant modes that can be originated in a
spherical structure. These functions $\phi_{k}\left(  \mathbf{r}\right)  $ are
expressed in a convenient spherical base and the frequencies of these modes
are given depending on the geometry. Hence, we get the following set of
differential equations
\begin{subequations}
\begin{align}
\ddot{q}_{0}(t)+\omega_{0}^{2}q_{0}(t)  &  =\sum_{k=1}^{\infty}c_{k}%
q_{k}(t),\label{subeq1}\\
\ddot{q}_{k}(t)+\omega_{k}^{2}q_{k}(t)  &  =c_{k}q_{0}(t) \label{subeq2}%
\end{align}
where, $\omega_{k}=\frac{\pi}{k}$ and $c_{k}=\frac{\omega_{k}}{R}\sqrt{2g}$.
The approximations considered in this model are associated with a very
specific and real situations. For instance, the object in the center of the
sphere can be considered a molecule or atom whose vibrational spectrum can be
approximated with a harmonic oscillator $q_{0}$.

Another situation which results in the same model is given by a small charge
distribution compared with the wavelengths of the incident radiation.
Using an appropriate gauge choice \cite{cohen,haroche}, the resulting model is
a system consisting of a harmonic oscillator coupled to the modes of a
radiation scalar field \cite{andion,Pimen1}.

As a third example, let us consider an ionized atom or molecule in a
general metalic cavity. The lagrangian operator for this system is%
\end{subequations}
\[
\hat{L}=\hat{L}_{mat}+\int_{V}d^{3}\mathbf{r}\left[  -\mathbf{\hat{E}}\left(
\mathbf{r},t\right)  \cdot\frac{1}{c}\frac{\partial}{\partial t}%
\mathbf{\hat{A}}\left(  \mathbf{r},t\right)  -\frac{\mathbf{\hat{E}}%
^{2}\left(  \mathbf{r},t\right)  +\mathbf{\hat{H}}^{2}\left(  \mathbf{r}%
,t\right)  }{2}+\frac{1}{c}\mathbf{\hat{\jmath}}\left(  \mathbf{r},t\right)
\cdot\mathbf{\hat{A}}\left(  \mathbf{r},t\right)  \right]
\]
where all products must be taken as symmetrized and $\hat{L}_{mat}$ describes
the harmonic oscillator $q_{0}$ and $\mathbf{\hat{\jmath}}\left(
\mathbf{r},t\right)  $ it is the electric current associated with $q_{0}$.

The field at the interior of the cavity must resonate in normal modes which
leads to a natural decomposition for the vector potential:%
\[
\int_{V}d^{3}\mathbf{rA}_{\alpha}\left(  \mathbf{r}\right)  \cdot
\mathbf{A}_{\beta}\left(  \mathbf{r}\right)  =\delta_{\alpha\beta}%
\]

\begin{align*}
\frac{1}{c}\mathbf{\hat{A}}\left(  \mathbf{r},t\right)   &  =\sum_{\alpha}%
\hat{q}_{\alpha}\left(  t\right)  \mathbf{A}_{\alpha}\left(  \mathbf{r}\right)
\\
-\mathbf{\hat{E}}\left(  \mathbf{r},t\right)   &  =\sum_{\alpha}\hat
{p}_{\alpha}\left(  t\right)  \mathbf{A}_{\alpha}\left(  \mathbf{r}\right) \\
\mathbf{\hat{\jmath}}\left(  \mathbf{r},t\right)   &  =\sum_{\beta}c_{\beta
}\hat{q}_{0}\left(  t\right)  \mathbf{A}_{\beta}\left(  \mathbf{r}\right)
\end{align*}

Using this together the radiation gauge and the boundary conditions, we find%
\begin{align*}
\hat{L}  &  =\hat{L}_{mat}+\left[  \sum_{\beta}\sum_{\alpha}\hat{p}_{\beta
}\left(  t\right)  \frac{d}{dt}\hat{q}_{\alpha}\left(  t\right)
\delta_{\alpha\beta}\right]  +\\
&  -\frac{1}{2}\left(  \left(  \sum_{\beta}\sum_{\gamma}\hat{p}_{\gamma
}\left(  t\right)  \hat{p}_{\beta}\left(  t\right)  \delta_{\gamma\beta
}\right)  \right)  +\\
&  -\frac{1}{2}\sum_{\beta}\sum_{\alpha}c^{2}\hat{q}_{\alpha}\left(  t\right)
\hat{q}_{\beta}\left(  t\right)  \int_{V}d^{3}\mathbf{rA}_{\beta}\left(
\mathbf{r}\right)  \cdot\left(  \mathbf{\nabla\times}\left(  \mathbf{\nabla
}\times\mathbf{A}_{\alpha}\left(  \mathbf{r}\right)  \right)  \right)  +\\
&  +\sum_{\beta}\hat{q}_{\beta}\left(  t\right)  c_{\beta}\hat{q}_{0}\left(
t\right)
\end{align*}

Choosing $\mathbf{A}_{\alpha}$\ as the eigenfunctions of the double-rot
operator\ ($\mathbf{\nabla\times}\left(  \mathbf{\nabla}\times\mathbf{\,}%
\right)  $):%
\[
\mathbf{\nabla\times}\left(  \mathbf{\nabla}\times\mathbf{A}_{\alpha}\left(
\mathbf{r}\right)  \right)  =\left(  \frac{\omega_{\alpha}}{c}\right)
^{2}\mathbf{A}_{\alpha}\left(  \mathbf{r}\right)
\]
we finally arrive at the very same model.

The only difference in the application of the model to each situation is the
spectrum $\omega_{\alpha}$ which will depend on the particular geometry of the
cavity. Naturally, it is possible to find many other physical systems
described by the very same model, but these three above are enough to
demonstrate its utility.

From now on let the symmetric Hamiltonian operator associated to this useful
model be%
\begin{align}
\hat{H}  &  =\frac{1}{2}\left\{  \hat{p}_{0}^{2}+\omega_{0}^{2}\hat{q}_{0}%
^{2}+\sum_{k=1}^{N}\left(  \hat{p}_{k}^{2}+\omega_{k}^{2}\hat{q}_{k}%
^{2}\right)  \right\} \nonumber\\
&  -\frac{1}{2}\sum_{k=1}^{N}c_{k}\left(  \hat{q}_{0}\hat{q}_{k}+\hat{q}%
_{k}\hat{q}_{0}\right)  , \label{eq:hamiltonian}%
\end{align}
in which the subindex $0$ is reserved to the oscillator in the center of the
sphere and the subindex $k=\left\{  1,...,N\right\}  $ is reserved to the
modes of the scalar field. It is simple to show that the equations
\eqref{subeq1} and \eqref{subeq2} can be derived from the Hamiltonian \eqref{eq:hamiltonian}.

The Hamiltonian can be written in a matrix form%
\begin{equation}
\hat{H}=\frac{1}{2}\hat{p}^{T}\hat{p}+\frac{1}{2}\hat{q}^{T}\Omega^{2}\hat{q}
\label{hamat}%
\end{equation}
defining the vectors%
\begin{subequations}
\begin{align}
\hat{p}^{T}  &  =\left(  \hat{p}_{0},\hat{p}_{1},...,\hat{p}_{k}\right)
,\label{eq:opervec}\\
\hat{q}^{T}  &  =\left(  \hat{q}_{0},\hat{q}_{1},...,\hat{q}_{k}\right)  .
\label{eq:2-1}%
\end{align}

The $\mathbf{\Omega}^{2}$ is a symmetrical interaction matrix given by%
\end{subequations}
\begin{equation}
\mathbf{\Omega}^{2}\mathbf{=}\left(
\begin{array}
[c]{ccccc}%
\omega_{0}^{2} & -c_{1} & -c_{2} & \ldots & -c_{N}\\
-c_{1} & \omega_{1}^{2} & 0 & \ldots & 0\\
-c_{2} & 0 & \omega_{2}^{2} & \ldots & 0\\
\vdots & \vdots & \vdots & \ddots & \vdots\\
-c_{N} & 0 & 0 & \ldots & \omega_{N}^{2}%
\end{array}
\right)  . \label{eq:matacoplamento}%
\end{equation}

The above matrix resumes all interactions in the system, those given by the
boundary conditions of the system associated with its geometry and manifested
in the frequencies of the field modes and those given by the dipolar interaction.

The matrix \eqref{eq:matacoplamento} can be diagonalized by an orthogonal
transformation $\mathbf{T}$,
\begin{equation}
\mathbf{D}^{2}=\mathbf{T\mathbf{\Omega}}^{2}\mathbf{T}^{T} \label{eq:ortran0}%
\end{equation}
with $\mathbf{D}^{2}=diag\left\{  \Omega_{0}^{2},\Omega_{1}^{2},\Omega_{2}%
^{2},...,\Omega_{N}^{2}\right\}  $ such that it is possible to express this
matrix as a product
\begin{equation}
\mathbf{D}^{2}=\mathbf{D}\cdot\mathbf{D}. \label{eq:ortran1}%
\end{equation}

The transformation \eqref{eq:ortran0} and the expression \eqref{eq:ortran1}
allow the factorization $\mathbf{\mathbf{\Omega}^{2}}=\mathbf{\Omega}%
\cdot\mathbf{\Omega}$,
\[
\mathbf{\Omega}\cdot\mathbf{\Omega}=\mathbf{TDT}^{T}\cdot\mathbf{TDT}^{T}%
\]
with $\mathbf{D}=diag\left\{  \Omega_{0},\Omega_{1},\Omega_{2},...,\Omega
_{N}\right\}  $ and
\begin{equation}
\mathbf{\Omega=TDT}^{T}. \label{eq:matsqt}%
\end{equation}
Actually, it is possible to find any power $\alpha$ of the $\Omega$ matrix,
\begin{equation}
\mathbf{\Omega^{\alpha}=TD^{\alpha}T}^{T}, \label{eq:matpot}%
\end{equation}
in which $\alpha\in\mathbb{R}$.

From now on, we assume each eigenvalue of $D^{\alpha}$ is such that
$\Omega_{i}^{\alpha}>0$. This excludes the case of complex frequencies which
could describe the penetration of each mode to the walls of the cavity and the
energy losses associated with it, but help to focus on the interaction of the
field with the dipole.

\section{Non-hermitean Operators}

The quadratic structure of the Hamiltonian \eqref{hamat} suggests, similarly
to an harmonic oscillator, that there is the possibility to define a set of
creation and annihilation operators for this system, meaning that the
Hamiltonian operator is factorable. Since in general such operators are not
connected with the particle content, Schwinger used to call them simply
\emph{non-hermitean operators}. In order to define them we have to use the
characteristics of the interaction matrix \eqref{eq:matacoplamento} and some
concepts originated in the theory of coupled harmonic oscillators at the classical level.

Similarly to the classical case, at the quantum level the state of motion of
each oscillator of the system depends on the state of motion of the other
elements of the system due to the interaction. In the same way, there are
states of motion of the system as a whole in which all of its constituents
moves with the same frequency and phase. Those states of motion constitute
the basic movements of the system, being called normal modes. There is one
normal mode for each element of the whole system.

In consequence, we are talking about two sets of operators: the first,
reporting the motions and excitations of each element of the system and the
second, reporting the motions and excitations of the system as a whole. In the
first set, each oscillator has a special behavior since its movements are
affected by the presence of other elements in the system. In the second set,
the group of elements of the system shows states of motion which have
associated a natural frequency called resonant frequency. Each combination of
motions is unique, with its own frequency, and all the combinations can be
considered as a system of uncoupled oscillators.

Given the characteristics of the two sets of operators mentioned above, we
shall call creation and annihilation operators of dressed states or DSO the
operators describing excitations of individual system components. Similarly,
the second set will be called annihilation and creation operators of the
normal modes or NMO. Thus, in the present section, we shall construct the two
sets of creation and annihilation operators and establish a simple
relationship between them.

\subsection{Dressed States Operators}

The annihilation and creation operators that we shall define are related with
the excitations of every one of the system components when the interaction is
considered. Thus, these operators will create a superposition of the
measurable states of each system component. The DSO and the states they create
are called dressed since just as it happens with the modification of the magnetic
moment of an atom by the coupling with a radio frequency field (see
\cite{CohenHaroche} section C) the frequencies and coupling constants of each
oscillator are affected by the interaction with the other elements of the
system. As a consequence, the energy states created by these operators are
different with respect to those which would present the components of the system
if there were no interaction.

On this way, following the structure of the matrices \eqref{eq:matpot}, the
DSO operators are given by the following expressions%

\begin{equation}%
\begin{array}
[c]{ccc}%
\hat{y} & = & \sqrt{\frac{1}{2\hbar}}\mathbf{\Omega}^{\frac{1}{2}}\left(
\hat{q}+i\mathbf{\Omega}^{-1}\hat{p}\right)  ,\\
\hat{y}^{T} & = & \sqrt{\frac{1}{2\hbar}}\left(  \hat{q}^{T}+i\hat{p}%
^{T}\mathbf{\Omega}^{-1}\right)  \mathbf{\Omega}^{\frac{1}{2}},\\
\hat{y}^{\dagger} & = & \sqrt{\frac{1}{2\hbar}}\mathbf{\Omega}^{\frac{1}{2}%
}\left(  \hat{q}-i\mathbf{\Omega}^{-1}\hat{p}\right)  ,\\
\hat{y}^{\dagger T} & = & \sqrt{\frac{1}{2\hbar}}\left(  \hat{q}^{T}-i\hat
{p}^{T}\mathbf{\Omega}^{-1}\right)  \mathbf{\Omega}^{\frac{1}{2}}.
\end{array}
\label{eq:opun}%
\end{equation}
where the operators $\hat{y}$, $\hat{y}^{\dagger}$and its transposed ones, are
vectors of the form%

\[%
\begin{array}
[c]{ccccc}%
\hat{y}^{T} & = & \left(
\begin{array}
[c]{cccc}%
\hat{y}_{0} & \hat{y}_{1} & ... & \hat{y}_{k}%
\end{array}
\right)  ,\text{and,} & \hat{y} & =\left(
\begin{array}
[c]{c}%
\hat{y}_{0}\\
\hat{y}_{1}\\
\vdots\\
\hat{y}_{k}%
\end{array}
\right)  ,
\end{array}
\]

\[%
\begin{array}
[c]{ccccc}%
\hat{y}^{\dagger T} & = & \left(
\begin{array}
[c]{cccc}%
\hat{y}_{0}^{\dagger} & \hat{y}_{1}^{\dagger} & ... & \hat{y}_{k}^{\dagger}%
\end{array}
\right)  ,\text{and,} & \hat{y}^{\dagger} & =\left(
\begin{array}
[c]{c}%
\hat{y}_{0}^{\dagger}\\
\hat{y}_{1}^{\dagger}\\
\vdots\\
\hat{y}_{k}^{\dagger}%
\end{array}
\right)  .
\end{array}
\]
As we can see, the transformations given by \eqref{eq:opun} keep the canonical
commutation relations
\begin{align}
\left[  \hat{y},\hat{y}^{\dagger}\right]   &  =\hat{y}^{T}\hat{y}^{\dagger
}-\hat{y}^{\dagger T}\hat{y}=\frac{1}{2\hbar}\left\{  -2i\hat{q}^{T}\hat
{p}+2i\hat{p}^{T}\hat{q}\right\} \nonumber\\
&  =\frac{i}{\hbar}\left\{  \hat{p}^{T}\hat{q}-\hat{q}^{T}\hat{p}\right\}  =1.
\label{eq:0}%
\end{align}
Therefore the transformation $\left(  \hat{q},\hat{p}\right)  \rightarrow
\left(  \hat{y},\hat{y}^{\dagger}\right)  $ is canonical and the inverse
relations of \eqref{eq:opun} are%

\begin{equation}%
\begin{array}
[c]{ccc}%
\hat{q} & = & \sqrt{\frac{\hbar}{2}}\mathbf{\Omega}^{-\frac{1}{2}}\left(
\hat{y}+\hat{y}^{\dagger}\right)  ,\\
\hat{p} & = & i\sqrt{\frac{\hbar}{2}}\mathbf{\Omega}^{\frac{1}{2}}\left(
\hat{y}^{\dagger}-\hat{y}\right)  ,\\
\hat{q}^{T} & = & \sqrt{\frac{\hbar}{2}}\left(  \hat{y}^{T}+\hat{y}^{\dagger
T}\right)  \mathbf{\Omega}^{-\frac{1}{2}},\\
\hat{p}^{T} & = & i\sqrt{\frac{\hbar}{2}}\left(  \hat{y}^{\dagger T}-\hat
{y}^{T}\right)  \mathbf{\Omega}^{\frac{1}{2}}.
\end{array}
\label{opdos}%
\end{equation}
Replacing \eqref{opdos} in the Hamiltonian \eqref{hamat} leads to the Dressed
State Hamiltonian:%

\begin{equation}
\hat{H}=\frac{\hbar}{2}\left\{  \hat{y}^{\dagger T}\mathbf{\Omega}\hat{y}%
+\hat{y}^{T}\mathbf{\Omega}\hat{y}^{\dagger}\right\}  . \label{hamontransform}%
\end{equation}
The solution of the problem implies the knowledge of the dynamics of each
system component. However, the fact that both the states of the oscillator as
the field modes are interdependent prevent us consider them isolated. This is
evidenced by the fact the states created by the DSO are in a complicated
entangled superposition. To solve this problem, we can make a transformation
from the DSO operators to a set of operators of the normal modes of the system
(NMO). This transformation can be performed using the same transformation
\textbf{$\mathbf{T}$} which was used in \eqref{eq:opun}. This transformation
allows to express the Hamiltonian of the system as the Hamiltonian of a set of
uncoupled free harmonic oscillators which can be solved in an exact way
allowing one to find the behavior of the matter oscillator and each one of the
modes of the field.

\subsection{Normal Modes Operators}

The Hamiltonian \eqref{hamontransform} is not in a diagonal form. Besides,
the matrix $\mathbf{\Omega}$ contains all the information about the
interactions and has a structure similar to a simple harmonic oscillator when
the Hamiltonian is in a matrix form, this implies that \eqref{hamontransform}
can actually be put in a diagonal form using $\mathbf{T}$ and the normal modes operators.

The normal modes $\hat{\xi}$ operators are defined as%

\begin{equation}%
\begin{array}
[c]{ccccccc}%
\hat{\xi} & = & T^{T}\hat{y} & ; & \hat{\xi}^{T} & = & \hat{y}^{T}T,\\
&  &  &  &  &  & \\
\hat{\xi}^{\dagger} & = & T^{T}\hat{y}^{\dagger} & ; & \hat{\xi}^{\dagger T} &
= & \hat{y}^{\dagger T}T,
\end{array}
\label{eq:trcoup}%
\end{equation}
in which%

\[%
\begin{array}
[c]{ccccc}%
\hat{\xi}^{T} & = & \left(
\begin{array}
[c]{cccc}%
\hat{\xi}_{0} & \hat{\xi}_{1} & ... & \hat{\xi}_{k}%
\end{array}
\right)  ,\text{and,} & \hat{\xi} & =\left(
\begin{array}
[c]{c}%
\hat{\xi}_{0}\\
\hat{\xi}_{1}\\
\vdots\\
\hat{\xi}_{k}%
\end{array}
\right)  ,
\end{array}
\]

\[%
\begin{array}
[c]{ccccc}%
\hat{\xi}^{\dagger T} & = & \left(
\begin{array}
[c]{cccc}%
\hat{\xi}_{0}^{\dagger} & \hat{\xi}_{1}^{\dagger} & ... & \hat{\xi}%
_{k}^{\dagger}%
\end{array}
\right)  ,\text{and,} & \hat{\xi}^{\dagger} & =\left(
\begin{array}
[c]{c}%
\hat{\xi}_{0}^{\dagger}\\
\hat{\xi}_{1}^{\dagger}\\
\vdots\\
\hat{\xi}_{k}^{\dagger}%
\end{array}
\right)  .
\end{array}
\]
Then, the Hamiltonian \eqref{hamontransform} can be written in its diagonal
form
\begin{align}
\hat{H}  &  =\frac{\hbar}{2}\left\{  \hat{\xi}^{\dagger T}\mathbf{D}\hat{\xi
}+\hat{\xi}^{T}\mathbf{D}\hat{\xi}^{\dagger}\right\} \label{hamonsuma}\\
&  =\sum_{k=0}^{N}\Omega_{k}\left(  \hat{\xi}_{k}^{\dagger}\hat{\xi}_{k}%
+\hat{\xi}_{k}\hat{\xi}_{k}^{\dagger}\right)  .\nonumber
\end{align}

Taking the relations $\left[  \hat{\xi}_{j},\hat{\xi}_{k}^{\dagger}\right]
=\delta_{jk}$, the last expression can be written%

\begin{equation}
\hat{H}=\hbar\sum_{k=0}^{N}\Omega_{k}\left(  \hat{\xi}_{k}^{\dagger}\hat{\xi
}_{k}+\frac{1}{2}\right)  =\hbar\sum_{k=0}^{N}\Omega_{k}\hat{\xi}_{k}%
^{\dagger}\hat{\xi}_{k}+\frac{\hbar}{2}Tr\mathbf{D}. \label{hamonsum}%
\end{equation}

The set of creation and annihilation operators of the normal modes create and
destroy the collective excitations of the system. In other words, it creates and
annihilates both the full field excitations and the excitations of the matter
oscillator. These collective excitations have a fixed number of energy quanta.
They are related to the collective behavior of the system and not with the
dynamics of each of its components at the individual level. Indeed, thanks to
the transformation $\mathbf{T}$ establishing a relationship between the
operators $(\hat{y}^{\dagger},\hat{y})$ and $(\hat{\xi}^{\dagger}\hat{\xi})$,
the dressed states of each system component can be expressed as a combination
of the normal mode states of the system and vice versa.

The normal modes are presented in equal number to the components of the system
and constitute a set of uncoupled harmonic oscillators. Each of these modes
has a natural frequency that is defined by the characteristics of the entire
system. Therefore, they are essentially a linear superposition of states of
motion for each system element. Also, since that the motion of each element of
the system is given by a unique combination of normal modes, we can infer that
their motion will not present a single frequency. Therefore, the quantum
states associated with each element of the system are not eigenstates of the
Hamiltonian of the whole system.

Each normal mode represents a way in which the energy is exchanged between the
elements of the system and it will depend on the number of \emph{quanta} that
is available between the field and the oscillator. Given that the number of
excitations is fixed, the amount of those excitations taken into account will
determine the size of the Hilbert space to be used. Also, since all of NMO are
essentially a set of uncoupled harmonic oscillators, the quantum states
associated with them can be represented as the tensor product of states
associated with each of these decoupled oscillators. The states associated to
DSO can be represented as a combination of the states associated to the NMO
but this combination is not a simple tensor product of the states since DSO
states are entangled.

\section{The Lagrangian and the Equations of Motion}

Given the Hamiltonian and taking into account that the system does not have
constrains, the inverse Legendre transform is
\[
\hat{L}=\frac{1}{2}\hat{p}^{T}\cdot\frac{d\hat{q}}{dt}+\frac{1}{2}\frac
{d\hat{q}^{T}}{dt}\cdot\hat{p}-\hat{H}.
\]
Using the transformations given in \eqref{opdos} we have for the Lagrangian of
the system in the DSO base the following canonical form
\begin{align}
\hat{L}  &  =i\frac{\hbar}{4}\left\{  \left(  \hat{y}^{\dagger T}-\hat{y}%
^{T}\right)  \left(  \frac{d\hat{y}}{dt}+\frac{d\hat{y}^{\dagger}}{dt}\right)
\right\} \nonumber\\
&  +i\frac{\hbar}{4}\left\{  \left(  \frac{d\hat{y}^{T}}{dt}+\frac{d\hat
{y}^{\dagger T}}{dt}\right)  \left(  \hat{y}^{\dagger}-\hat{y}\right)
\right\}  -\hat{H}\nonumber\\
&  =\frac{i\hbar}{4}\left\{  \hat{y}^{\dagger T}\frac{d\hat{y}}{dt}-\hat
{y}^{T}\frac{d\hat{y}^{\dagger}}{dt}\right\} \nonumber\\
&  -\frac{\hbar}{2}\left\{  \hat{y}^{\dagger T}\mathbf{\Omega}\hat{y}+\hat
{y}^{T}\mathbf{\Omega}\hat{y}^{\dagger}\right\}  . \label{ledest1}%
\end{align}

In the same way, with the transformation made in \eqref{eq:trcoup} the
Lagrangian in the NMO base is
\begin{align}
\hat{L}  &  =\frac{\hbar}{2}\left\{  i\hat{\xi}^{\dagger T}\frac{d\hat{\xi}%
}{dt}-i\hat{\xi}^{T}\frac{d\hat{\xi}^{\dagger}}{dt}-\left(  \hat{\xi}^{\dagger
T}\mathbf{D}\hat{\xi}+\hat{\xi}^{T}\mathbf{D}\hat{\xi}^{\dagger}\right)
\right\} \nonumber\\
&  =\frac{\hbar}{2}\sum_{k=0}^{N}\left\{  i\hat{\xi}_{k}^{\dagger}\frac
{d\hat{\xi}_{k}}{dt}-i\hat{\xi}_{k}\frac{d\hat{\xi}_{k}^{\dagger}}{dt}%
-\Omega_{k}\left(  \hat{\xi}_{k}^{\dagger}\hat{\xi}_{k}+\hat{\xi}_{k}\hat{\xi
}_{k}^{\dagger}\right)  \right\}  . \label{eq:qulag}%
\end{align}
In the last expression it has been neglected the terms related with the total
derivative, since they just generate the transformation from DSO base to the
NMO one.

\subsection{Equations of Motion}

Schwinger Action Principle \cite{schwingercd} can be used for the derivation
of equations of motion of the quantum operators. We choose to perform a
variation on the quantum Lagrangian \eqref{eq:qulag} in the NMO description
since in this base the matrix equations are diagonal and decoupled and the
expression of any quantity emerged from here is suitable to be manipulated.
After some operations we have
\begin{align}
\delta\hat{L}  &  =\frac{\hbar}{2}\delta\hat{\xi}^{\dagger T}\left(
i\frac{d\hat{\xi}}{dt}-\mathbf{D}\hat{\xi}\right)  -\frac{\hbar}{2}\hat{\xi
}^{T}\mathbf{D}\delta\hat{\xi}^{\dagger}\nonumber\\
&  -\delta\hat{\xi}^{T}\left(  i\frac{d\hat{\xi}^{\dagger}}{dt}+\mathbf{D}%
\hat{\xi}^{\dagger}\right)  -\frac{\hbar}{2}\hat{\xi}^{\dagger T}%
\mathbf{D}\delta\hat{\xi}\nonumber\\
&  +\frac{i\hbar}{2}\hat{\xi}^{\dagger T}\frac{d\left(  \delta\hat{\xi
}\right)  }{dt}-\frac{i\hbar}{2}\hat{\xi}^{T}\frac{d\left(  \delta\hat{\xi
}^{\dagger}\right)  }{dt}, \label{eq:varequlag}%
\end{align}
then, taking $\delta\hat{L}=0,$ we obtain%
\[%
\begin{array}
[c]{ccc}%
i\frac{d\hat{\xi}}{dt}-\mathbf{D}\hat{\xi}=0 & \text{, and} & i\frac{d\hat
{\xi}^{\dagger}}{dt}+\mathbf{D}\hat{\xi}^{\dagger}=0,\\
&  & \\
i\frac{d\hat{\xi}^{T}}{dt}-\hat{\xi}^{T}\mathbf{D}=0 & \text{, and} &
i\frac{d\hat{\xi}^{\dagger T}}{dt}-\hat{\xi}^{\dagger T}\mathbf{D}=0.
\end{array}
\]
The solutions for the last matrix differential equations are given by%
\begin{align}
\hat{\xi}(t)  &  =\exp\left[  -i\mathbf{D}t\right]  \hat{\xi}\left(
t_{0}\right) \label{eq:solutions}\\
&
\mbox{, and \ensuremath{\hat{\xi}^{\dagger}}(t)=\ensuremath{\exp\left[i\mathbf{D}t\right]\hat{\xi}^{\dagger}\left(t_{0}\right)},}
\label{solutum}\\
\hat{\xi}^{T}(t)  &  =\hat{\xi}^{T}\left(  t_{0}\right)  \exp\left[
-i\mathbf{D}t\right] \nonumber\\
&  \mbox{, and }\hat{\xi}^{\dagger T}(t)=\hat{\xi}^{\dagger T}\left(
t_{0}\right)  \exp\left[  i\mathbf{D}t\right]  ,\nonumber
\end{align}
in which the vectors $\hat{\xi}(t_{0})$, $\hat{\xi}^{\dagger}(t_{0})$,
$\hat{\xi}^{T}(t_{0})$ e $\hat{\xi}^{\dagger T}(t_{0})$ are defined as the set
of operators with initial conditions, $t=t_{0}$.

\subsection{Generalized Transformation Function}

Following Schwinger's Action Principle, the operator $\hat{G}(t)$ is the
generator of the admissible infinitesimal transformations for the system. Its
functional form can be obtained taking the solutions of the equations
\eqref{eq:qulag} and the Hamiltonian \eqref{hamonsum}, then
\begin{align*}
\delta\left\langle \xi^{\dagger},t|\xi,t_{0}\right\rangle  &  =\left\langle
\xi^{\dagger},t\right\vert \hat{G}\left(  t\right)  -\hat{G}\left(
t_{0}\right)  \left\vert \xi,t_{0}\right\rangle \\
&  =\left\langle \xi^{\dagger},t\right\vert \left.  \left(  \hat{p}\delta
\hat{q}-\hat{H}\delta t\right)  \right\vert _{t_{0}}^{t_{1}}\left\vert
\xi,t_{0}\right\rangle .
\end{align*}
Taking the relations (\ref{eq:opun}) and the Hamiltonian (\ref{hamontransform}%
) we obtain the explicity form of the generator $\hat{G}\left(  t\right)  $ in
the DSO base:%
\begin{equation}
\hat{G}=i\frac{\hbar}{2}\left(  \hat{y}^{\dagger T}-\hat{y}^{T}\right)
\left(  \delta\hat{y}+\delta\hat{y}^{\dagger}\right)  -\frac{\hbar}{2}\left\{
\hat{y}^{\dagger T}\mathbf{\Omega}\hat{y}+\hat{y}^{T}\mathbf{\Omega}\hat
{y}^{\dagger}\right\}  \delta t. \label{eq:actprinca}%
\end{equation}
Similarly, using the transformations (\ref{eq:trcoup}) in the last expression,
we obtain the same generator in the NMO base:
\begin{equation}
\hat{G}=\hbar\sum_{k=0}^{N}\left\{  \frac{i}{2}\left(  \hat{\xi}_{k}^{\dagger
}\delta\hat{\xi}_{k}-\hat{\xi}_{k}\delta\hat{\xi}_{k}^{\dagger}\right)
-\Omega_{k}\left(  \hat{\xi}_{k}^{\dagger}\hat{\xi}_{k}+\frac{1}{2}\right)
\delta t\right\}  . \label{eq:actprinc-1}%
\end{equation}

Since the expressions related to \eqref{eq:actprinc-1} are decoupled, we can
take the solutions \eqref{eq:solutions} for a generic component $k$ and
calculate the transition amplitude $\left\langle \xi^{\dagger},t|\xi
,t_{0}\right\rangle $. In order to do that, we need to solve the variational
differential equation (\ref{eq:actprinc}), in which we substitute
(\ref{eq:solutions}) such that $\delta\hat{\mathcal{W}}=\hat{G}\left(
t\right)  -\hat{G}\left(  t_{0}\right)  $ is given by%
\begin{align*}
\delta\hat{\mathcal{W}}  &  =\hbar\sum_{k=0}^{N}\left\{  \frac{i}{2}\left(
\hat{\xi}_{k}^{\dagger}\delta\hat{\xi}_{k}\left(  t_{0}\right)  e^{-i\Omega
_{k}t}-\hat{\xi}_{k}\left(  t_{0}\right)  \delta\hat{\xi}_{k}^{\dagger
}e^{-i\Omega_{k}t}\right)  \right. \\
&  \left.  -\Omega_{k}\left(  \hat{\xi}_{k}^{\dagger}\hat{\xi}_{k}+\frac{1}%
{2}\right)  \delta t\right\}  .
\end{align*}

In this way, taking $\left[  \hat{\xi}_{k}^{\dagger},\delta\hat{\xi}%
_{k}\right]  =\left[  \delta\hat{\xi}_{k}^{\dagger},\hat{\xi}_{k}\right]  =0$,
which are the commutation relations derived from the Schwinger's Action
Principle \cite{schwingercd}, one finds
\[
\delta\hat{\mathcal{W}}=-\hbar\sum_{k=0}^{N}\left\{  \frac{i}{2}\delta\left\{
\hat{\xi}_{k}^{\dagger}\hat{\xi}_{k}\left(  t_{0}\right)  \right\}
e^{-i\Omega_{k}t}-\left(  \hat{\xi}_{k}^{\dagger}\hat{\xi}_{k}+\frac{1}%
{2}\right)  \Omega_{k}\delta t\right\}  .
\]
Performing an integration of the variational equation \eqref{eq:actprinc},
we have%
\begin{align*}
\hat{\mathcal{W}}  &  =-i\hbar\sum_{k=0}^{N}\xi_{k}^{\dagger}\xi_{k}\left(
t_{0}\right)  e^{-i\Omega_{k}t}-\frac{\hbar}{2}\sum_{k=0}^{N}\Omega_{k}t\\
&  =-i\hbar\sum_{k=0}^{N}\xi_{k}^{\dagger}\xi_{k}\left(  t_{0}\right)
e^{-i\Omega_{k}t}-\frac{\hbar}{2}Tr\mathbf{D}t,
\end{align*}
and using (\ref{amplitude}), leave a transformation function
\begin{align}
\langle\xi^{\dagger},t_{1}|\xi,t_{0}\rangle &  =\exp\left\{  -\frac{i}%
{2}Tr\mathbf{D}t+\sum_{k=0}^{N}\xi_{k}^{\dagger}\xi_{k}\left(  t_{0}\right)
e^{-i\Omega_{k}t}\right\} \nonumber\\
&  =\exp\left\{  -\frac{i}{2}Tr\mathbf{D}t\right\}  \exp\left\{  \vec{\xi}%
{}^{\dagger T}e^{-i\mathbf{D}t}\vec{\xi}\left(  t_{0}\right)  \right\}  .
\label{eq:TrasFunc}%
\end{align}

\section{Energy Spectrum}

In the NMO base, the Hamiltonian of the system \eqref{hamonsum} is totally
diagonal and equivalent to a set of $N$ intedependent harmonic oscillators,
each of them is actually a collective motion of the entire system. Given that
the transformation $\mathbf{T}$ at the quantum level represents a canonical
transformation, the spectrum of the system in the NMO base should be totally
equivalent to the spectrum of the Hamiltonian \eqref{eq:hamiltonian}. As
noted above, the Hamiltonian in the NMO base is diagonal and its Hilbert space
can be constructed as the product Hilbert space of $N$ independent harmonic
oscillators \cite{berezin} such that
\begin{equation}
\hat{H}=\sum_{k=0}^{N}\hat{H}_{k}=\hbar\sum_{k=0}^{N}\Omega_{k}\left(
\hat{\xi}_{k}^{\dagger}\hat{\xi}_{k}+\frac{1}{2}\right)  , \label{eq:se1}%
\end{equation}
in which the Hilbert space of each one is given by the following tensor
product
\begin{equation}
\hat{H}_{k}=\hat{1}_{0}\otimes\hat{1}_{1}\otimes...\otimes\hat{1}_{k-1}%
\otimes\hat{h}_{k}\otimes\hat{1}_{k+1}\otimes...\otimes\hat{1}_{N},
\label{eq:se2}%
\end{equation}
where $\hat{h}_{k}$ is the hamiltonian of one single normal mode. The identity
of the \emph{kth} Hilbert space is given as
\begin{equation}
\hat{1}_{k}=\sum_{n_{k}}\left\vert n_{k}\right\rangle \left\langle
n_{k}\right\vert .
\end{equation}
On this way the total identity for the system is
\begin{align}
\hat{1}  &  =\prod\limits_{i=0}^{N}\otimes\hat{1}_{i}=\prod\limits_{i=0}%
^{N}\otimes\sum_{n_{i}}\left\vert n_{i}\right\rangle \left\langle
n_{i}\right\vert \nonumber\\
&  =\prod\limits_{i=0}^{N}\sum_{n_{i}}\left\vert n_{i}\right\rangle
\left\langle n_{i}\right\vert . \label{eq:se3}%
\end{align}

A generic element of the set of states can be given by
\begin{equation}
\left\vert N\right\rangle =\left\vert n_{0},n_{1},...,n_{N}\right\rangle
=\left\vert n_{0}\right\rangle \otimes\left\vert n_{1}\right\rangle
\otimes...\otimes\left\vert n_{N}\right\rangle , \label{eq:se4}%
\end{equation}
and the action of the Hamiltonian over this eigenstates leaves
\begin{align*}
\hat{H}\left\vert n_{0},n_{1},...,n_{N}\right\rangle  &  =\sum_{k=0}^{N}%
\hat{H}_{k}\left\vert n_{0},n_{1},...,n_{N}\right\rangle \\
&  =\sum_{k=0}^{N}\left\vert n_{0}\right\rangle \otimes...\otimes\hat{h}%
_{k}\left\vert n_{k}\right\rangle \otimes...\otimes\left\vert n_{N}%
\right\rangle \\
&  =\sum_{k=0}^{N}E_{n_{k}}\left\vert n_{0},n_{1},...,n_{N}\right\rangle .
\end{align*}
It means that the energy spectrum is just a simple sum:
\begin{equation}
E_{n_{0}n_{1}...n_{N}}=\sum_{k=0}^{N}E_{n_{k}}. \label{eq:se6}%
\end{equation}

Now we can calculate the transition amplitude for the system in the initial
arbitrary state
\begin{equation}
\left\vert \xi,t_{0}\right\rangle =\prod\limits_{k=0}^{N}\otimes\left\vert
\xi_{k},t_{0}\right\rangle =\left\vert \xi_{0},\xi_{1},...\xi_{k},...\xi
_{N},t_{0}\right\rangle . \label{eq:se7}%
\end{equation}
Using the relations (\ref{eq:se3}) and the orthogonality relation $\langle
n_{i}|m_{j}\rangle=\delta_{m,m;i,j}$, we have
\begin{align*}
\left\vert \xi,t\right\rangle  &  =e^{-i\frac{\hat{H}}{\hbar}t}\left\vert
\xi,t_{0}\right\rangle \\
&  =\exp\left\{  -i\frac{t}{\hbar}\sum_{k=0}^{N}\hat{H}_{k}\right\}
\left\vert \xi_{0},\xi_{1},...\xi_{k},...\xi_{N},t_{0}\right\rangle \\
&  =\exp\left\{  -i\frac{t}{\hbar}\sum_{k=0}^{N}E_{k}\right\}  \prod
\limits_{k=0}^{N}\left\langle n_{k}|\xi_{k},t_{0}\right\rangle .
\end{align*}
Therefore,%
\begin{equation}
\left\langle \xi^{\dagger},t_{1}|\xi,t_{0}\right\rangle =\prod\limits_{j=0}%
^{N}\left(  \sum_{n_{j}}e^{-\frac{i}{\hbar}E_{n_{j}}t}\left\vert \left\langle
n_{j}|\xi_{j},t_{0}\right\rangle \right\vert ^{2}\right)  . \label{eq:es8}%
\end{equation}
On the other hand, due to the series expansion of the transformation function
(\ref{eq:TrasFunc}),
\begin{equation}
\left\langle \xi^{\dagger},t_{1}|\xi,t_{0}\right\rangle =\prod\limits_{k=0}%
^{N}\sum_{n=0}^{\infty}\frac{\xi_{k}^{\dagger n}\xi_{k}^{n}\left(
t_{0}\right)  }{n!}e^{-i\Omega_{k}\left(  n+\frac{1}{2}\right)  t}.
\label{eq:es9}%
\end{equation}
Comparing \eqref{eq:es8} and \eqref{eq:es9},%
\[
\prod\limits_{k=0}^{N}\left(  \sum_{n=0}^{\infty}e^{-\frac{i}{\hbar}E_{n_{k}%
}t}\left\vert \left\langle n_{k}|\xi_{k},t_{0}\right\rangle \right\vert
^{2}\right)  =\prod\limits_{k=0}^{N}\sum_{n=0}^{\infty}\frac{\xi_{k}^{\dagger
n}\xi_{k}^{n}\left(  t_{0}\right)  }{n!}e^{-i\Omega_{k}\left(  n+\frac{1}%
{2}\right)  t}.
\]
we obtain the eigenfunctions
\[
\left\langle \xi_{k}^{\dagger},t_{0}|n_{k}\right\rangle =\frac{\xi
_{k}^{\dagger n}}{\sqrt{n!}}\text{, and }\left\langle n_{k}|\xi_{k}%
,t_{0}\right\rangle =\frac{\xi_{k}^{n}}{\sqrt{n!}},
\]
and the exact form of the spectrum of the system
\begin{align*}
E_{n_{0}n_{1}...n_{N}}  &  =\sum_{k=0}^{N}E_{n_{k}}=\sum_{k=0}^{N}\hbar
\Omega_{k}\left(  n_{k}+\frac{1}{2}\right) \\
&  =\sum_{k=0}^{N}\hbar\Omega_{k}n_{k}+\frac{\hbar}{2}Tr\mathbf{D.}%
\end{align*}
From the last expression for the spectrum, we have the vacuum energy of the
system
\[
E_{\underset{\text{N-times}}{\underbrace{0,0,0,0...0}}}=\frac{\hbar}%
{2}Tr\mathbf{D}=\frac{\hbar}{2}Tr\mathbf{\mathbf{\Omega}}.
\]

\section{Transition Probabilities}

The expression for the transition amplitude \eqref{eq:TrasFunc} is given in
the NMO base of the system. However, this information is shown in a useless
form if we want to study the behavior of an individual element of the system
since the states associated with NMO are a superposition of DSO.

The transition amplitude in the NMO base can be understood in the following
way. Let us take $|N\rangle$ and $|M\rangle$ two arbitrary states in the
normal base, then
\begin{align*}
\left\langle M\right\vert e^{-i\frac{\hat{H}}{\hbar}t}\left\vert
N\right\rangle  &  =\prod\limits_{k=0}^{N}e^{-i\hbar\Omega_{k}\left(
n_{k}+\frac{1}{2}\right)  t}\delta_{m_{k},n_{k}}\\
&  =e^{-i\hbar\sum_{s=0}^{N}\Omega_{s}\left(  n_{s}+\frac{1}{2}\right)
t}\prod\limits_{k=0}^{N}\delta_{m_{k},n_{k}}.
\end{align*}

This expression is the probability amplitude that the system in the state
$|N\rangle$ at the initial time do a transition to the state $|M\rangle$ at
the time $t$. However, to extract useful physical information as, for
instance, the behavior of the oscillator in the center of the cavity or the
dynamics of a specific field mode, we need to evaluate the transition
amplitude between the two corresponding dressed states of the system. This
information is achieved in the non-diagonalized basis. For this reason we
reformulate the generalized transition amplitude \eqref{eq:es9} to the DSO
representation using the transformation \eqref{eq:trcoup}:%
\begin{equation}
\xi_{k}^{\dagger}=\sum_{j=0}^{N}\left[  T\right]  _{jk}y_{j}^{\dagger
},\label{nmotodso}%
\end{equation}
in which $\left[  T\right]  _{jk}$ is a element of the $\mathbf{T}$ matrix and
the $\xi_{k}^{\dagger}$ is the eigenvalue of the operator $\hat{\xi}%
_{k}^{\dagger}$ acting over the $|\xi_{k}\rangle$, associated to the the
$k$-th normal mode. Thus, using \eqref{nmotodso}, the expression
(\ref{eq:es9}) is given by%
\begin{equation}
\langle y^{\dagger},t|y,t_{0}\rangle=\exp\left\{  -\frac{i}{2}Tr\mathbf{D}%
t\right\}  \prod_{r=0}^{N}\prod_{s=0}^{N}\exp\left\{  y_{s}^{\dagger
}\mathcal{J}_{rs}\left(  t\right)  y_{r}\left(  t_{0}\right)  \right\}
,\label{elementtransic}%
\end{equation}
in which it is defined
\begin{equation}
\sum_{k=0}^{N}\left[  T\right]  _{sk}\left[  T\right]  _{kr}e^{-i\Omega_{k}%
t}=\mathcal{J}_{rs}\left(  t\right)  .\label{element}%
\end{equation}
The expression \eqref{elementtransic} is the generalized transition
amplitude for the system in the Dressed State representation. After some
manipulations, we can extract information about the transition between any
two dressed modes both of the field or the oscillator in the center of the
cavity:%
\begin{equation}
\langle y_{s}^{\dagger},t|y_{r},t_{0}\rangle=\exp\left\{  -\frac{i}%
{2N}Tr\mathbf{D}t\right\}  \exp\left\{  y_{s}^{\dagger}\mathcal{J}_{rs}\left(
t\right)  y_{r}\left(  t_{0}\right)  \right\}  .\label{ampdsoind}%
\end{equation}
The above expression can also be decomposed in the number base for the dressed
states, using
\begin{equation}
|\langle y_{k}^{\dagger},t_{0}|\overline{n}_{k}\rangle=\frac{y_{k}%
^{\dagger\overline{n}_{k}}}{\sqrt{\overline{n_{k}}!}}\quad\text{ and}%
\quad|\langle\overline{m}_{k}|y_{k},t_{0}\rangle=\frac{y_{k}^{\overline{m_{k}%
}}}{\sqrt{\overline{m_{k}}!}},\label{eq:transform1}%
\end{equation}
in which the overline means that the number state is associated with the
dressed base. Then, the transition \eqref{ampdsoind} between any two states
in the DSO base is given by%
\begin{equation}
\langle y_{s}^{\dagger},t|y_{r},t_{0}\rangle=\exp\left\{  -\frac{i}%
{2N}Tr\mathbf{D}t\right\}  \sum_{\overline{n}=0}^{\infty}\frac{y_{s}%
^{\dagger\overline{n}}\left(  t\right)  }{\sqrt{\overline{n}!}}\mathcal{J}%
_{rs}^{\overline{n}}\left(  t\right)  \frac{y_{r}^{\overline{n}}\left(
t_{0}\right)  }{\sqrt{\overline{n}!}},\label{eq:tran1}%
\end{equation}
in which the subscripts $r$ and $s$ are related with the individual components
of the system. From the expressions \eqref{eq:transform1} and
\eqref{eq:tran1} we obtain the following transition amplitude
\[
\left\langle n_{s},t|n_{r}\right\rangle =\exp\left\{  -\frac{i}{2N}%
Tr\mathbf{D}t\right\}  \mathcal{J}_{rs}^{n}\left(  t\right)  ,
\]
\[
\mathcal{J}_{rs}^{n}\left(  t\right)  =\sum_{\substack{0\leq l_{m}\leq
n\\\sum_{j=0}^{m}l_{j}}}^{n}\frac{n!}{l_{0}!l_{1}!l_{2}!...l_{m}!}%
\prod\limits_{p=0}^{m}\left(  \left[  T\right]  _{sk}\left[  T\right]
_{kr}e^{-i\Omega_{k}t}\right)  ^{l_{p}}.
\]
For the probabilities, we can show that
\[
\sum_{r=0}^{N}\left\vert \mathcal{J}_{rs}\left(  t\right)  \right\vert ^{2}=1,
\]
knowing that $\mathbf{T}$ is orthogonal. Writting this expression for $n=0$
corresponding to the oscillator and in the center of the cavity and
$k=\left\{  1,...,N\right\}  $ for the modes of the field, one finds
\[%
\begin{array}
[c]{cc}%
\left\vert \mathcal{J}_{00}\left(  t\right)  \right\vert ^{2}+\sum_{k=0}%
^{N}\left\vert \mathcal{J}_{0k}\left(  t\right)  \right\vert ^{2} & =1,\\
\left\vert \mathcal{J}_{k_{1}0}\left(  t\right)  \right\vert ^{2}+\sum
_{k_{2}=0}^{N}\left\vert \mathcal{J}_{k_{1}k_{2}}\left(  t\right)  \right\vert
^{2} & =1.
\end{array}
\]
Such expressions are related with the probabilities to have a transition
between any two elements of the system with $n$ \emph{quanta}. Then, the atom
or any field mode initially in the $n$-th state can change to any another
state in a later time \cite{andion,Pimen1}.

\section{Conclusions and Outlook}

In the case of interacting systems modeled by (\ref{subeq1}) and
(\ref{subeq2}), we show that it is possible to define unambiguously a
transformation in which some analogies with a classical coupled systems can be
considered. Here, the system, as a whole can be studied as a set of coupled
harmonic oscillators each of them having a different coupling constant. With
the proposed approach, the normal modes operators create simple decoupled
states associated with the whole system. On the other hand, the dressed states
operators create quantum states that each individual oscillator can show in
the presence of interaction with the rest of the system.

In the \emph{Dressed State} formalism introduced in
\cite{cohen,polonsky,haroche}, the states of each element of the system are
constructed by hand, expressing them as a special combination of the states of
the whole system, such construction leads to entangled states. Besides that,
in this work, we have introduced those states by mean of the \emph{Dressed
State Operators} (DSO) which create those \emph{Dressed States} naturally. In
our approach those states are not constructed by hand, it is the structure of
the DSO which leads to the entangled states, as it can be easily checked solving
the problem for only one mode of the field.

The quantum state of each part of the system, the oscillator or each field
mode, is defined with a set of parameters which is modified by the presence
of the interaction. For instance, the frequency is redefined as a function of
the other parameters of the system
\[
\omega\rightarrow\Omega\left(  \omega_{0},\left\{  \omega_{j}\right\}
,\left\{  c_{k}\right\}  ;j,k\in\left\{  1,2,...,N\right\}  \right)  ,
\]
if they are compared with the free interaction parameters, $c_{i}=0$ in \eqref{eq:hamiltonian}.

The orthogonality of the transformation $\mathbf{T}$ preserves the geometrical
structure of the operators, being at the quantum level a canonical
transformation which preserves the canonical structure of the theory in both,
normal and dressed representations. Thus, the transformation functions in the
normal and dressed representations have the same information. In the dressed
state representation, the transformation function allows us to know all the
amplitude transitions between the states of each one of the elements in the
system since the \ dressed states are not eigenstates of the system.

The transformations \eqref{eq:opun}, \eqref{opdos} and \eqref{eq:trcoup}
show that in some cases the Rotating Wave Approximation (RWA) is unnecessary.
This is because those transformations lead to a system which can be solved in
an exact way, what means that the RWA approximation it is not necessary if
the correct definition of the transformation is used.

The Dressed State Approach, developed in this work, can be applied to the
study of problems in which the dissipation and the interaction of a system with a
bath or reservoir can be also included. This approach can even be suitable to
deal with non-linear interacting system via a perturbative treatment. Some of
these points will be explored elsewhere.

\section{Acknowledgments}

B.M.Pimentel thanks CNPq for partial support and J.A.Ramirez thanks
CAPES for full support. J.A.Ramirez would like to thank to the Institute for
Theoretical Physics at the S\~{a}o Paulo State University in which part of the
work was developed.

\end{document}